\documentclass[11pt]{article}
\usepackage{moriond,epsfig}

\def\NPB{{\em Nucl. Phys.} B}
\def\PLB{{\em Phys. Lett.}  B}
\def\PRL{{\em Phys. Rev. Lett.}}
\def\PRD{{\em Phys. Rev.} D}

\def\be{\begin{equation}}
\def\ee{\end{equation}}
\def\bea{\begin{eqnarray}}
\def\eea{\end{eqnarray}}

\def\dz{D\O}
\def\ipb{\ensuremath{\mathrm{pb}^{-1}}}
\def\bp{\bar{p}}
\def\SM{Standard Model}
\def\pt{\ensuremath{p_T}}
\def\z{\ensuremath{\mathrm{Z}}}
\def\w{\ensuremath{\mathrm{W}^\pm}}
\def\met{\hbox{{\rm $E_T$}\hskip-1.2em{/}}}

\begin{document}
\vspace*{4cm}
\title{ELECTROWEAK MEASUREMENTS AT TEVATRON}

\author{ FR\'ED\'ERIC D\'ELIOT \\
(On behalf of the CDF and \dz\ collaborations)}

\address{CEA-Salcay, DSM/DAPNIA/SPP, bat 141\\
91191 Gif-sur-Yvette Cedex, France}

\maketitle\abstracts{
Recent Run~2 experimental results on electroweak physics from the CDF and \dz\ collaborations at 
Tevatron are reviewed. We present measurements of the inclusive W and Z boson cross-sections times
leptonic branching ratio as well as the $\z/\gamma^*$ forward-backward asymmetry and the \w\ 
charge asymmetry for $p \bp$ collisions at $\sqrt{s}=1.96$~TeV .
The first Run~2 studies for the W~width and W~mass measurements are also reported.
The di-boson production results are presented elsewhere~\cite{gilles}.
}

\section{Introduction}
The Run 2 of the Tevatron is well underway and the accelerator, which collides protons against anti-protons at a 
center-of-mass energy of 1.96~TeV had delivered at the beginning of 2005 around $0.8~fb^{-1}$ of integrated luminosity 
to the two experiments CDF and \dz. CDF and \dz\ are analyzing these data to precisely study the properties of the \w\ 
and \z\ bosons performing an extensive electroweak program. 
This program starts by the measurements of the single W and Z production 
cross-sections with the boson decaying leptonically which allows to test \SM\ predictions.
The $\tau$ decay channel also allows to demonstrate the ability of the experiments to identify $\tau$'s. 

Measuring forward-backward $\z/\gamma \to e^- e^+$ asymmetry provides further tests of the \SM, while
the \w\ charge asymmetry measurement provides constraints on the parton distribution functions (PDFs) of the proton. 
Another important part of the Tevatron electroweak program consists in measuring the W~width and mass since a precise 
determination of the W~mass together with the top mass measurement leads to an indirect determination of the 
mass of the still undiscovered Higgs boson.

Studying diboson production also gives constraints on
new physics beyond the \SM. The Tevatron results on this subject is discussed elsewhere~\cite{gilles}.

\section{W and Z cross-sections measurements}
\label{sec:xsec}

\subsection{W and Z cross-sections in electron(s) and muon(s) decay channels}

The W and Z productions provide clean, abundant and well known signals. These signals both 
allow to test the \SM\ by performing cross-section ratios and are extensively used to calibrate 
and understand the detectors response as well as to measure identification efficiencies.
The precise measurements of these cross-sections can also provide an independent
cross-check of the luminosity measurement mainly determined by the total rate of inelastic $p \bp$ collisions.
At the Tevatron, single W and Z bosons are predominantly produced in the s-channel via quark anti-quark fusion.
Due to the high level of QCD background, the signal from bosons decaying in the electron or muon channel is 
easier to identify.

\begin{figure}[!htb]
\begin{center}
\psfig{figure=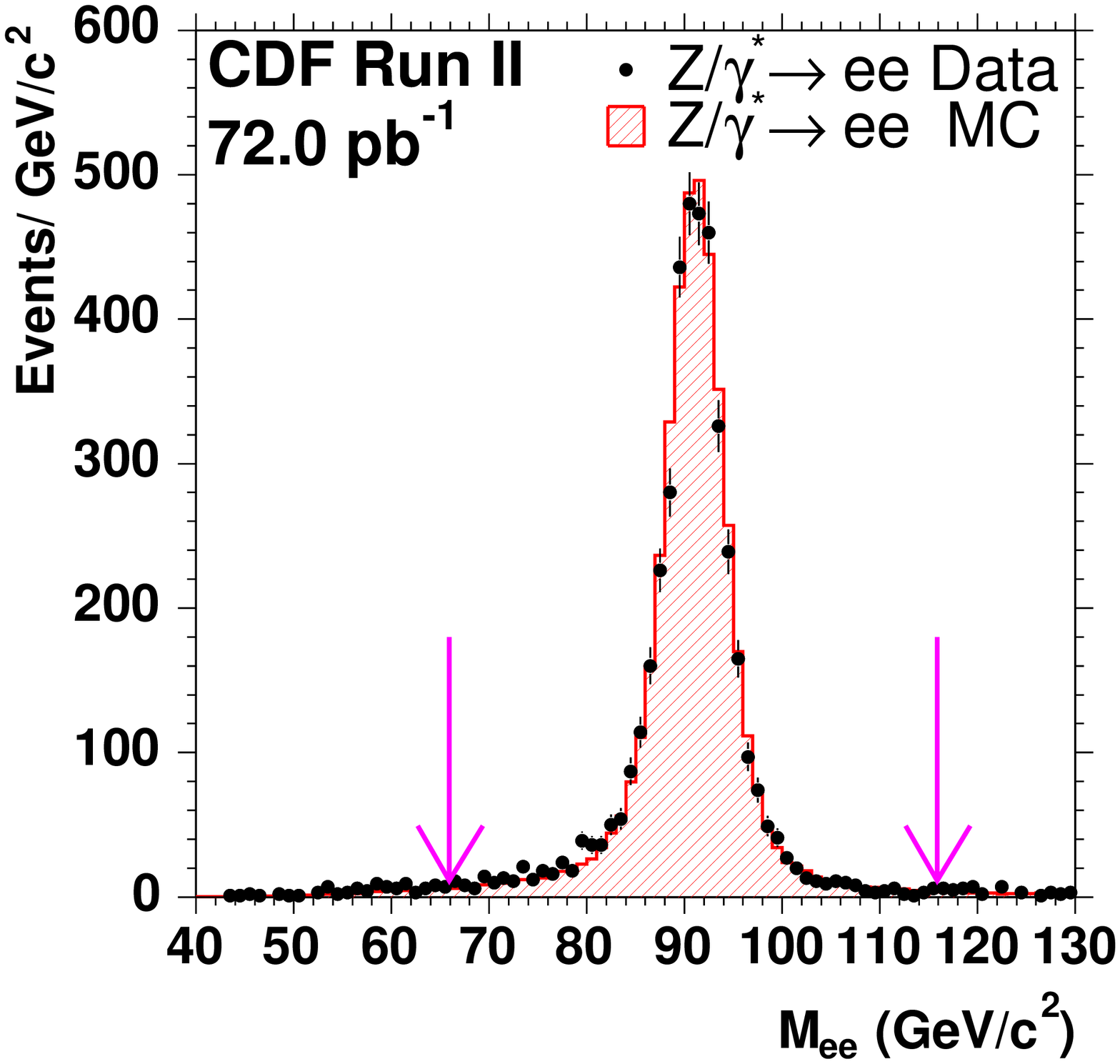,width=5cm}
\psfig{figure=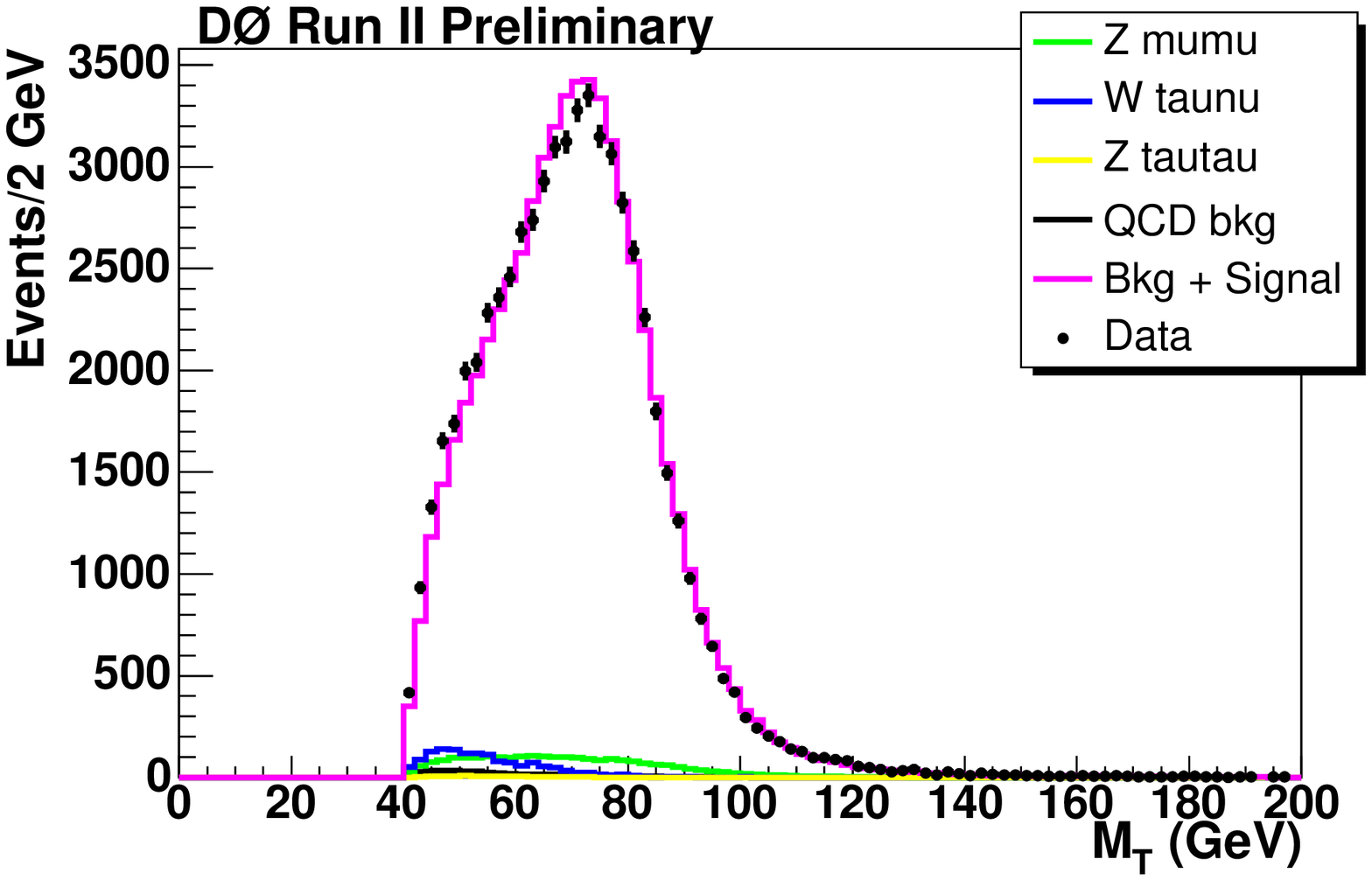,width=7.5cm}
\caption{CDF invariant mass distribution for $\z \to e^-e^+$ candidates (left) and 
\dz\ transverse mass distribution for $W \to \mu\nu$ candidates (right). The W transverse mass is defined as: 
$M_T = \sqrt{2 p_{T \ell} \met~~(1 - cos(\phi_\ell - \phi_{\met}~~))}$~where $\met~~$ is the 
missing transverse energy, $\phi_\ell$ the azimuthal direction of this missing energy, \pt\ the transverse 
momentum of the charged lepton and $\phi_\ell$ its azimuthal angle. 
\label{fig:wz_kinem}}
\end{center}
\end{figure}

The $\w \to \ell \nu$ and $\z \to \ell^+ \ell^-$ signals are characterized by at least one high transverse momentum 
charged lepton ($\pt >$~15 to 25~GeV). In addition, the \z\ decay leads to a second high \pt\ charged lepton while the \w\ decay 
generates large missing transverse energy ($\met~>$~15 to 25~GeV) due to the undetected neutrino.
Figure~\ref{fig:wz_kinem} shows an example of \z\ invariant mass and \w\ transverse mass distributions.

The main sources of background for those signals are QCD dijet events (where either the jets are faking or emitting a charged lepton or 
where jets are faking missing transverse energy) and electroweak background (W and Z decaying into 
$\tau$(s) and $Z \to \ell \ell$ for the W signal when one lepton escapes detection). 
The QCD background is evaluated using data and represents 1 to 2\% while the electroweak background is evaluated using Monte Carlo 
simulation (1 to 6~\%).

Tables~\ref{tab:cdf_emu_xsec} and \ref{tab:d0_emu_xsec} summarize CDF and \dz\ results. It can be seen that the largest error 
comes from the luminosity determination (around 6\%). Apart from this one, the cross-section measurements are limited by
systematic uncertainties. The main source of systematic uncertainties comes from the PDFs error (around 1.5\%).

\begin{table}[!htb]
\caption{Summary of the \w\ and \z\ cross-section times branching ratio measurements at CDF Run~2.\label{tab:cdf_emu_xsec}}
\begin{center}
\begin{tabular}{|c|c|c|}
\hline
Process & Integrated luminosity used & $\sigma$.Br in pb measured in CDF \\
\hline
$Z \to ee$     & $72~\ipb$ & $255.8 \pm 3.9 (stat) \pm 5.5 (sys) \pm 15 (lumi)$~\cite{cdf_xsec} \\
\hline
$Z \to \mu\mu$ & $72~\ipb$ & $248 \pm 5.9 (stat) \pm 7.6 (sys) \pm 15 (lumi) $~\cite{cdf_xsec} \\
\hline
$W \to e\nu$   & $72~\ipb$ & $2780 \pm 14 (stat) \pm 60 (sys) \pm 166 (lumi) $ (central)~\cite{cdf_xsec} \\
               & $64~\ipb$ & $2874 \pm 34 (stat) \pm 167 (sys) \pm 172 (lumi)$ (plug) \\
\hline
$W \to \mu\nu$ & $72~\ipb$ & $2768 \pm 16 (stat) \pm 64 (sys) \pm 166 (lumi) $~\cite{cdf_xsec} \\

\hline
\end{tabular}
\end{center}
\end{table}

\begin{table}[!htb]
\caption{Summary of the \w\ and \z\ cross-section times branching ratio measurements at \dz\ Run~2.\label{tab:d0_emu_xsec}}
\begin{center}
\begin{tabular}{|c|c|c|}
\hline
Process & Integrated luminosity used & $\sigma$.Br in pb measured in \dz \\
\hline
$Z \to ee$     & $177~\ipb$ & $264.9 \pm 3.9 (stat) \pm 9.9 (sys) \pm 17.2 (lumi)$\\
\hline
$Z \to \mu\mu$ & $148~\ipb$ & $291.3 \pm 3.0 (stat) \pm 6.9 (sys) \pm 18.9 (lumi)$\\
\hline
$W \to e\nu$   & $177~\ipb$ & $2865 \pm 8.3 (stat) \pm 76 (sys) \pm 186 (lumi)$ \\
\hline
$W \to \mu\nu$ & $96~\ipb$  & $2989 \pm 15 (stat) \pm 81 (sys) \pm 194 (lumi)$ \\
\hline
\end{tabular}
\end{center}
\end{table}

Using these results, one can compute the R ratio of the W to Z cross-sections in the electron or muon channel 
(see equation~\ref{eq:R}) where the luminosity uncertainties cancel.
CDF reports the following R ratio for the combined electron and muon channel using 72~\ipb:
$$R(e+\mu) = 10.92 \pm 0.15 (stat) \pm 0.14 (sys).$$
\dz\ reports a R ratio in the electron channel using 177~\ipb:
$$R(e) = 10.82 \pm 0.16 (stat) \pm 0.28 (sys).$$
This ratio is related to the total W width $\Gamma_W^{tot}$ in the following way:
\begin{equation}
R = \frac{\sigma(p \bp \to W) \times Br(W \to \ell \nu)}{\sigma(p \bp \to Z) \times Br(Z \to \ell^+ \ell^-)}
= \frac{\sigma(W)}{\sigma(Z)} \frac{1}{Br(Z \to \ell^+ \ell^-)} \frac{\Gamma(W \to \ell \nu)}{\Gamma_W^{tot}}.
\label{eq:R}
\end{equation}
Taking the NLO calculation~\cite{ratio_wz_calc} for the ratio of the production cross-sections $\frac{\sigma(W)}{\sigma(Z)}$, 
the Z to charged lepton branching ratio $Br(Z \to \ell \ell)$ from the world average~\cite{pdg} (dominated by LEP measurements) and 
the theoritical value for the W partial width into lepton~\cite{pdg} $\Gamma(W \to \ell \nu)$, one can extract 
the W total width $\Gamma_W^{tot}$. CDF made this extraction and reported $\Gamma_W^{tot} = 2.079 \pm 0.041$~GeV compatible with
the world average~\cite{pdg04} $\Gamma_W^{tot} = 2.124 \pm 0.041$~GeV and the \SM\ expectation~\cite{pdg04}
$\Gamma_W^{tot} = 2.092 \pm 0.003$~GeV.

\subsection{\w\ and \z\ cross-sections into tau(s)}
In a hadronic environment, the reconstruction of $\tau$ leptons is challenging. Measuring the W and Z cross-sections 
in the $\tau$ channel proves the ability of the detector to reconstruct $\tau$'s in addition to provide test of the \SM.
Final states into $\tau$ leptons are also interesting as they are predicted and sometimes even enhanced in models of new physics.

\dz\ has measured the $\z \to \tau^+ \tau^-$ cross-section with one $\tau$ decaying into a muon ($\tau \to \mu \nu_\mu \nu_\tau$) 
on which one can trigger and the other one into hadrons~\cite{d0_ztau}. A neural network technique is used to identify the
$\tau$'s decaying hadronically depending of their decay products (single $\pi^\pm$, multiple $\pi^\pm$'s or $\rho$ decay).
The main backgrounds for this channel arise from QCD events (around 50\%) and $\w \to \mu \nu$ or $\z \to \mu \mu$ (around 6\%).
The main systematic uncertainties come from trigger efficiency determination (3.5\%) and QCD background
determination performed on data (3.5\%). The result using $226~\ipb$ of data is:
\begin{equation}
\sigma(p \bp \to Z) \times Br(Z \to \tau \tau) = 237 \pm 15 (stat) \pm 18 (sys) \pm 15 (lumi)~pb~(\dz).
\label{eq:ztau}
\end{equation}

CDF has measured the $W \to \tau \nu$ cross-section when the $\tau$ is decaying hadronically. The $\tau$ is reconstructed 
as one or three charged tracks that match a narrow calorimeter cluster.
The main backgrounds for this channel come again from QCD dijet events (15\%) and from $W \to e \nu$ (4\%).
The main source of systematic uncertainty is the $\tau$ identification efficiency determination (5.5\%).
Using $72~\ipb$ of data, the result is:
\begin{equation}
\sigma(p \bp \to W) \times Br(W \to \tau \nu) = 2620 \pm 70 (stat) \pm 21 (sys) \pm 160 (lumi)~pb~(CDF).
\label{eq:wtau}
\end{equation}

\subsection{Cross-section summary}
\begin{figure}[!htb]
\begin{center}
\psfig{figure=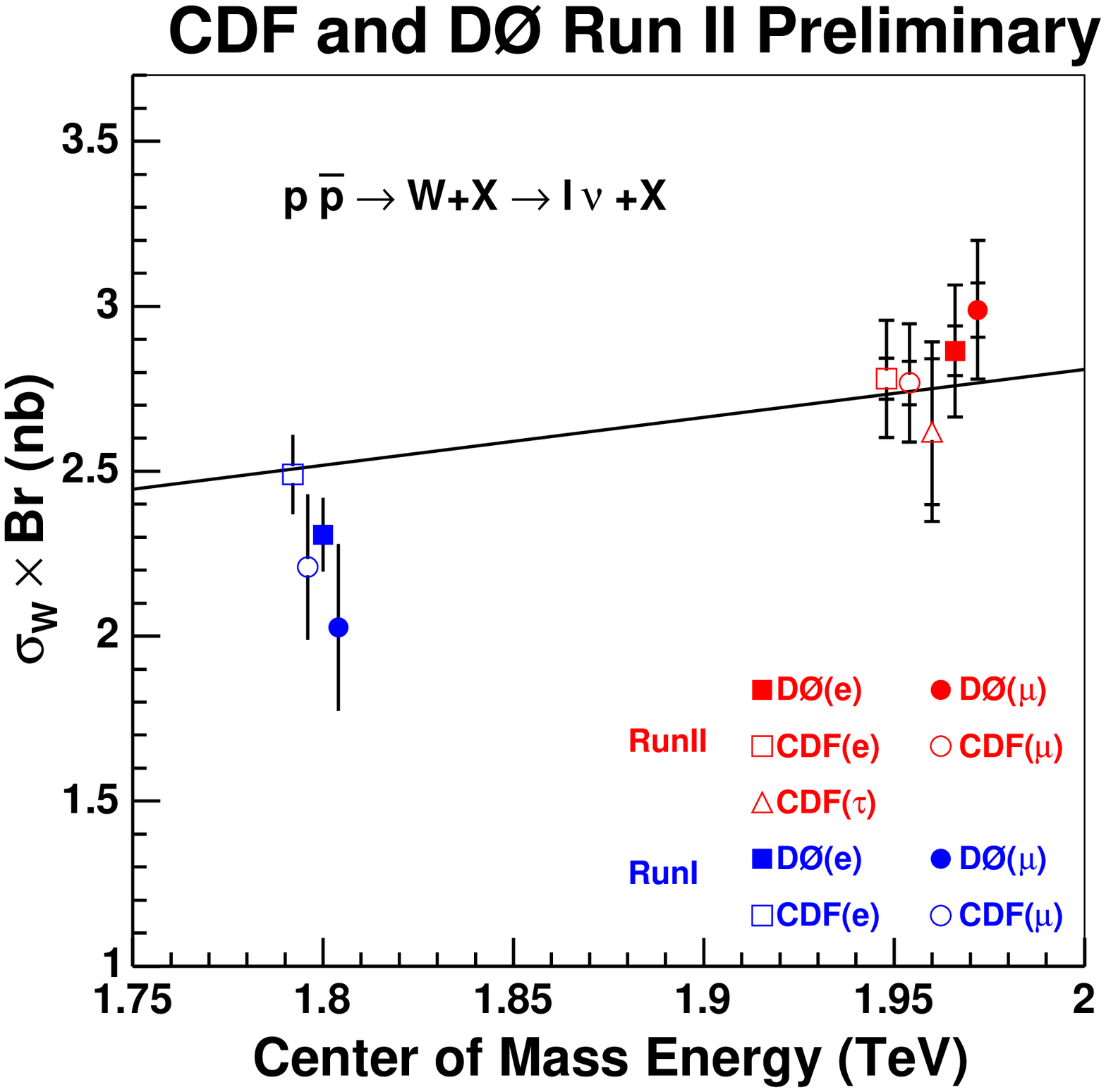,width=7.cm}
\psfig{figure=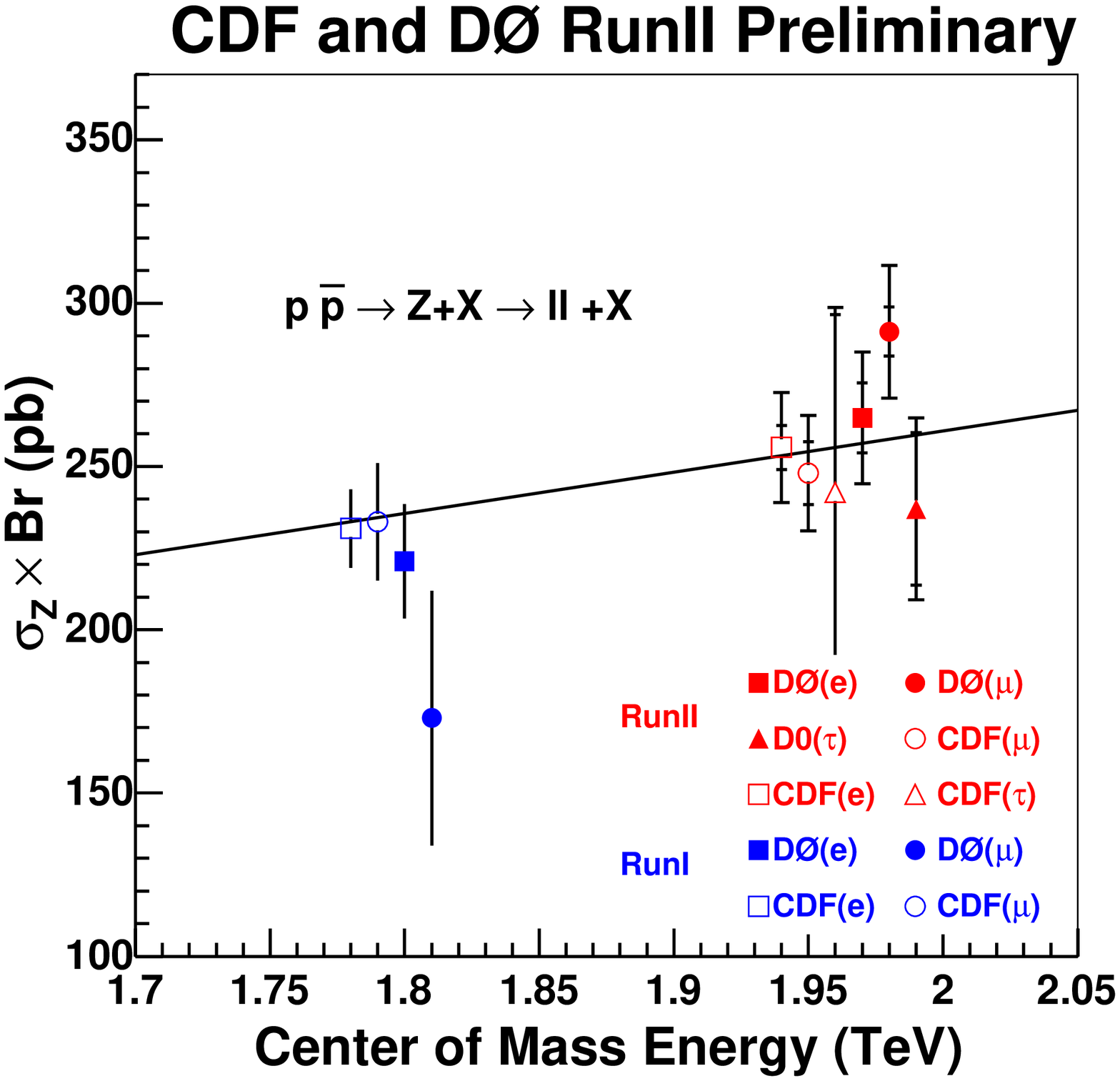,width=7.cm}
\caption{Summary of the W and Z cross-section measurements at Tevatron. The inner most error bars 
combine only the statistical and systematical errors while the outer error bars correspond to the total error 
including the luminosity error. The curve is the \SM\ expectation \protect \cite{SM_xsec}.
\label{fig:wz_xsec}}
\end{center}
\end{figure}

All the Tevatron inclusive W and Z cross-section measurements are summarized in Figure~\ref{fig:wz_xsec}.

The ratio of the $W \to \mu \nu$ and $W \to e \nu$ cross-sections or the ratio of the 
$W \to \tau \nu$ and $W \to e \nu$ cross-sections provide test of the lepton coupling universality 
to the W boson. CDF has for instance reported a $e-\tau$ ratio that is well compatible with lepton universality:
$$\frac{g_{\tau}}{g_e} = 0.99 \pm 0.02 (stat) \pm 0.04 (sys).$$

\section{$\z/\gamma^*$ forward-backward asymmetry}
The vector and axial-vector nature of the fermion couplings to the \z\ leads to an asymmetry in
the \z\ decay lepton angle in the process $p \bp \to \z/\gamma^* \to \ell^+ \ell^-$.
If $\theta$ is the angle between the incoming quark and the outgoing lepton $\ell^-$ in the 
rest frame of the dilepton pair, the dilepton production cross-section can be written as:
\begin{equation}
\frac{d \sigma(qq' \to \ell \ell)}{d \cos \theta} = \alpha \left( \beta(1 + \cos^{2} \theta) + A_{FB} \cos \theta \right)
\label{eq:ll_xsec}
\end{equation}
where $\alpha$ and $\beta$ are constant and $A_{FB}$ is the $\z/\gamma^*$ forward-backward asymmetry which can
be defined as:
\begin{equation}
A_{FB} = \frac{N_F - N_B}{N_F + N_B} = 
	\frac{\sigma_{\cos \theta > 0} - \sigma_{\cos \theta < 0}}{\sigma_{\cos \theta > 0} + \sigma_{\cos \theta < 0}}.
\label{eq:afb}
\end{equation}

This asymmetry can be measured by counting the normalized difference between the number of events 
with positive ($N_F$) and negative ($N_B$) $\cos \theta$. Besides the ability to probe the relative strength of 
couplings between the \z\ boson and the quarks, 
$A_{FB}$ is also sensitive to additional non-SM contribution to the $p \bp \to \ell^+ \ell^-$ process that are predicted by Z' or 
extra-dimension models for instance. Measuring $A_{FB}$ at the Tevatron is then complementary to direct searches in the high dilepton 
invariant mass region above LEP sensitivity.

\begin{figure}[!htb]
\begin{center}
\psfig{figure=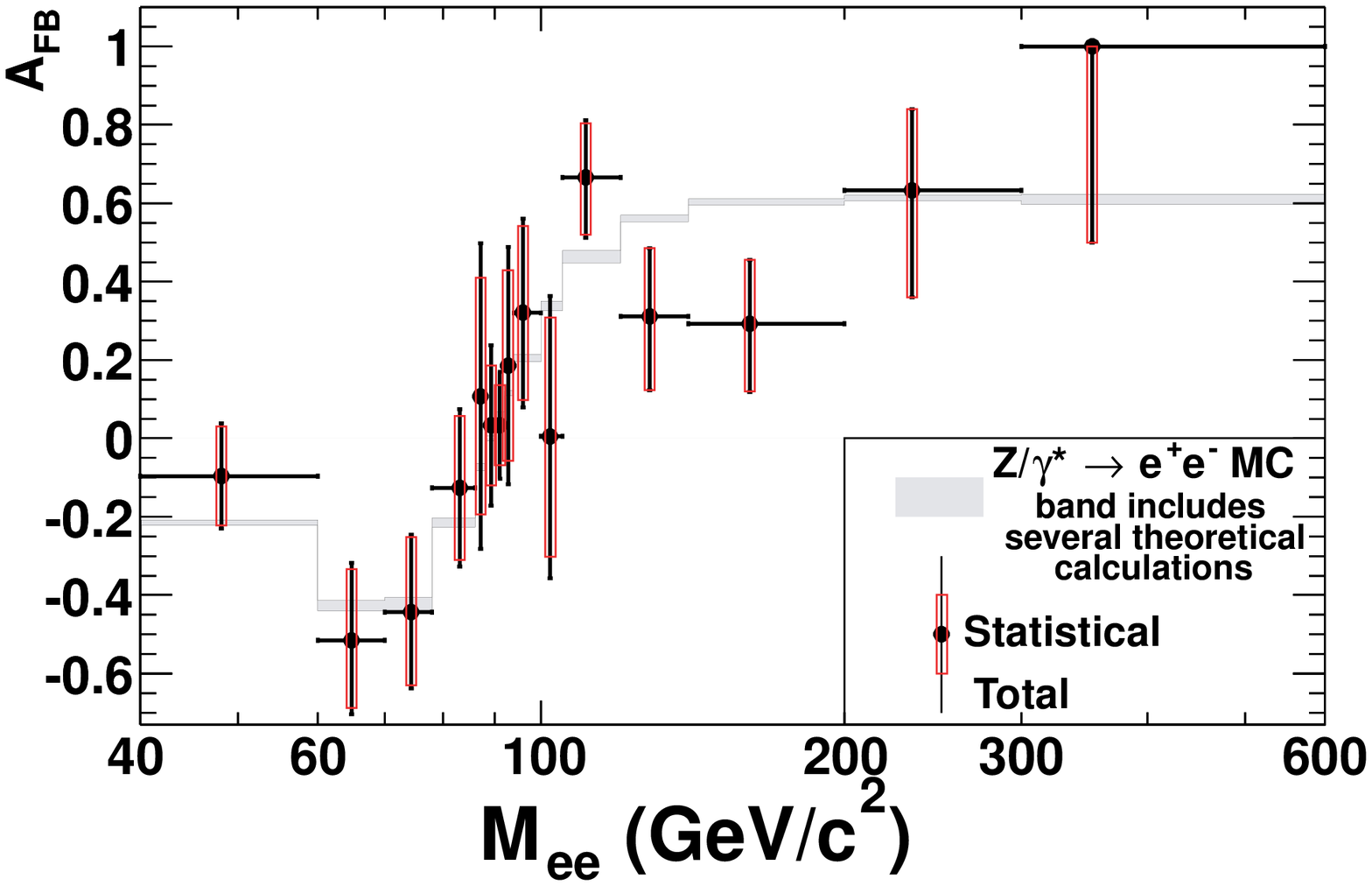,width=7.5cm}
\psfig{figure=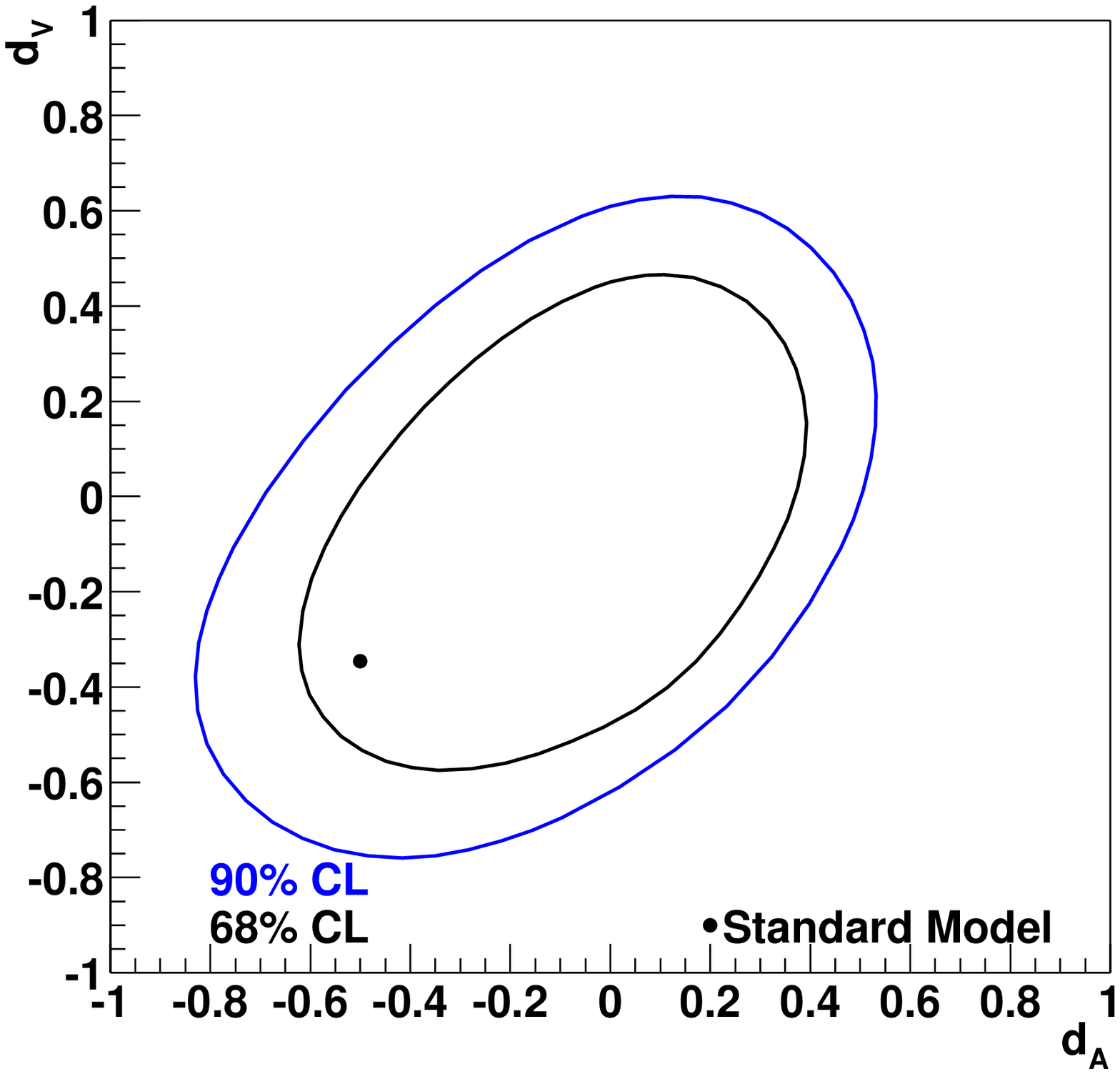,width=6cm}
\caption{Left: CDF $A_{FB}$ measurement with statistical and systematical uncertainties (crosses) and SM prediction (histogram).
Right: CDF 68\% and 90\% confidence level contours for the d~quark couplings to the \z\ boson. The dot is the SM prediction.
\label{fig:afb}}
\end{center}
\end{figure}

CDF made the measurement using $72~\ipb$ selecting high \pt\ dielectrons both in the central and plug (forward) 
calorimeters~\cite{cdf_afb}.
The main background in the sample comes from dijet events (around 3\%). An unconstrained unfolding method is used to take into
account the acceptance and the $M_{ee}$ bin migration effects due to radiation and detector energy resolution.
The result of the $A_{FB}$ measurement is shown in Figure~\ref{fig:afb} (left).
Expressing $A_{FB}$ in term of \z-quark and \z-electron coupling constants, one can extract the values of these
couplings from the $A_{FB}$ measurement. The results are shown in Figure~\ref{fig:afb} (right) for the d~quark couplings. 
Finally performing a fit where the quark and electron couplings to the \z\ are expressed as a function of $\sin^2 \theta^{eff}_W$ gives
$\sin^2 \theta^{eff}_W = 0.2238 \pm 0.0040 (stat) \pm 0.0030 (sys)$ with $\chi^2/ndf = 12.5/14.0$. These results are compatible
with the SM predictions but are currently statistically limited. They will improve with higher statistics.

\section{\w\ charge asymmetry}
As we have seen for the cross-section measurements (section~\ref{sec:xsec}), many of the precision
measurements at a hadron collider are limited by the PDFs uncertainties. Improving the knowledge of
these PDFs is then an important task. One way to achieve this is to measure the \w\ charge asymmetry 
which then gives an input on the u and d components of the proton.

As the u~quark carries on average a higher fraction of the proton momentum than the d~quark, the 
$W^+$ produced via the process $u \bar{d} \to W^+$ tends to be boosted in the proton direction (u direction).
This leads to a forward-backward charge asymmetry for the \w\ bosons defined as:
\begin{equation}
A(y_W) = \frac{d \sigma(W^+) / dy_W - d \sigma(W^-) / dy_W}
	{d \sigma(W^+) / dy_W + d \sigma(W^-) / dy_W}
\label{eq:ayw}
\end{equation}
where $y$ is the rapidity of the W boson and $d \sigma(\w) / dy_W$ the \w\ differential
production cross-sections. The W rapidity $y_W$ is sensitive to the difference
between the u and d quark momenta. As it is not possible to measure directly the rapidity 
of the W bosons due to the unknown longitudinal momentum of the escaping neutrino from the W decay,
one instead measures the pseudo-rapidity of the decay lepton $\eta_{\ell}$.

CDF reports a measurement of the \w\ charge asymmetry $A(\eta_e)$ in the electron channel~\cite{cdf_ayw} using $170~\ipb$.
Here the main experimental challenge is to measure the charge misidentification in the forward region
i.e. to understand the tracking system in the forward region where the PDFs are less constrained. The main background in 
this channel comes from dijet events (2\% in the central part and 15\% in the forward calorimeter).

\begin{figure}[!htb]
\begin{center}
\psfig{figure=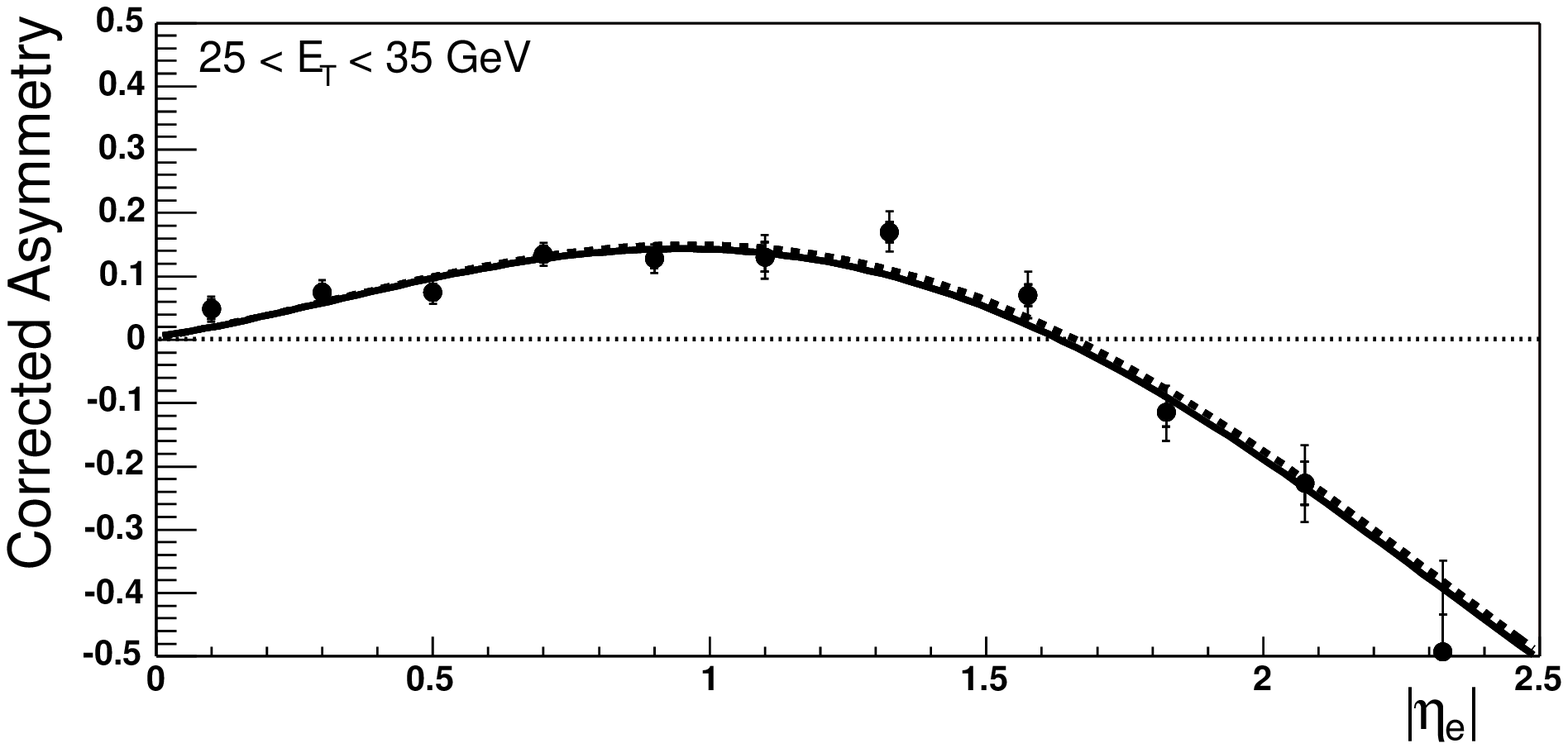,width=7.5cm}
\psfig{figure=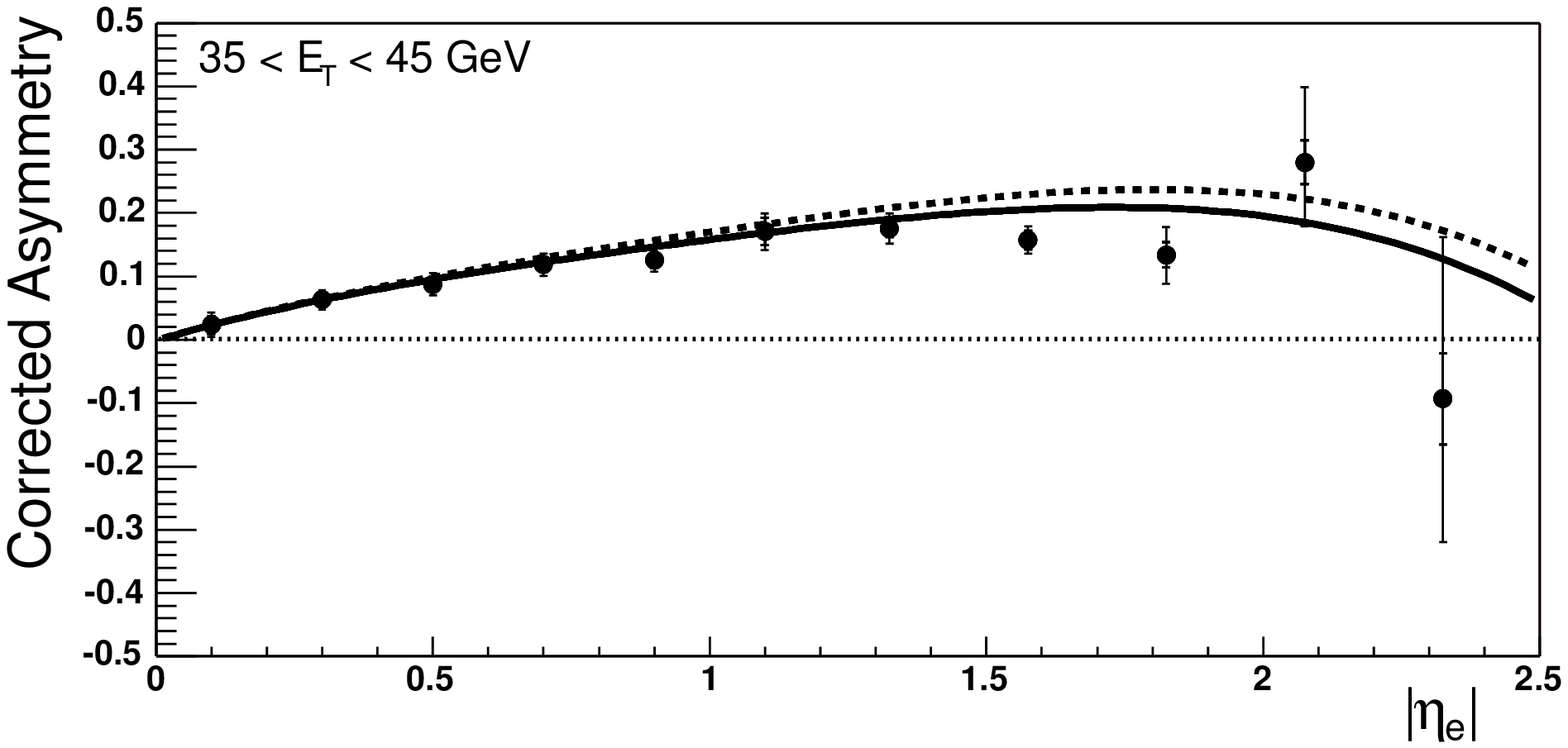,width=7.5cm}
\caption{CDF Measured W charge asymmetry in 2 different $E_T$ bins (crosses) compared to the predictions 
using 2 different PDFs and the NLO soft gluon resummation programm RESBOS (plain line: CTEQ6.1M~\protect\cite{cteq}
+ RESBOS~\protect\cite{resbos}, dashed line: MRST02~\protect\cite{mrst} + RESBOS~\protect\cite{resbos}).
\label{fig:ayw}}
\end{center}
\end{figure}

The results are shown in Figure~\ref{fig:ayw} after correction for the charge misidentification and after background subtraction. 
As no indication of CP asymmetry between positive and negative $\eta_e$ values is found, the results
are shown as a function of $|\eta_e|$. The shape of the curves comes from two competing effects: first the boost of the $W^+$
in the proton direction as explained above and second the V-A nature of the $e^+-W^+$ coupling that favors
a positron emission opposite to the $W^+$ direction in the $W^+$ rest frame. Future inclusion of these results in the PDF fits
will allow to reduce PDFs uncertainties.

\section{\w\ width and mass}
Precise measurements of the W~mass and couplings test the nature of the electroweak symmetry breaking.
The top, W and Higgs masses are actually related through the following formula:
\begin{equation}
M_W^2 = \frac{\pi \alpha(M_Z^2)}{\sqrt{2}G_F} \frac{1}{1 - M_W^2/M_Z^2} \frac{1}{1 - \Delta r}
\label{eq:mw}
\end{equation}
where the electroweak loop corrections are hidden in $\Delta r$ 
($\alpha$ is the electromagnetic coupling constant and $G_F$ the Fermi coupling constant measured in muon decay).
$\Delta r$ depends on the top and Higgs masses in the following way $\Delta r \sim M_t^2$ and $\Delta r \sim ln (M_H)$.
So measuring $M_W$ and $M_t$ yields indirect constraints on the Higgs mass~\cite{lepewwg}.

As the longitudinal momentum of neutrino in the process $W \to \ell \nu$ is not known, the 
W~mass (resp. width) is extracted from the Jacobian edge (resp. high mass tail) of the 
W transverse mass defined as $M_T = \sqrt{2 p_{T \ell} p_{T \nu} (1 - \cos(\phi_\ell - \phi_\nu)) }$.
These measurements require the best possible knowledge of the detector reponses.

\begin{figure}[!htb]
\begin{center}
\psfig{figure=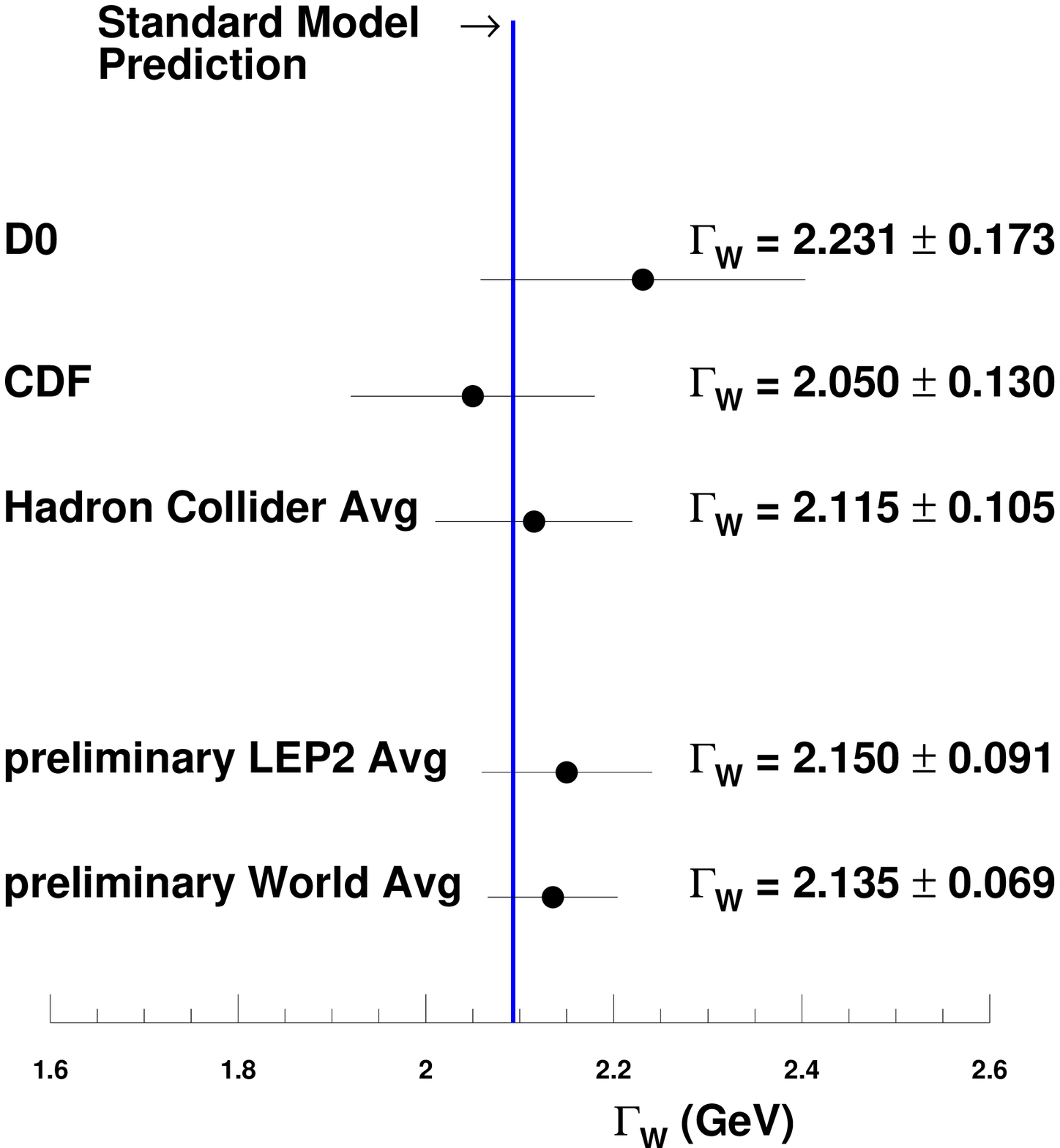,width=7.cm}
\psfig{figure=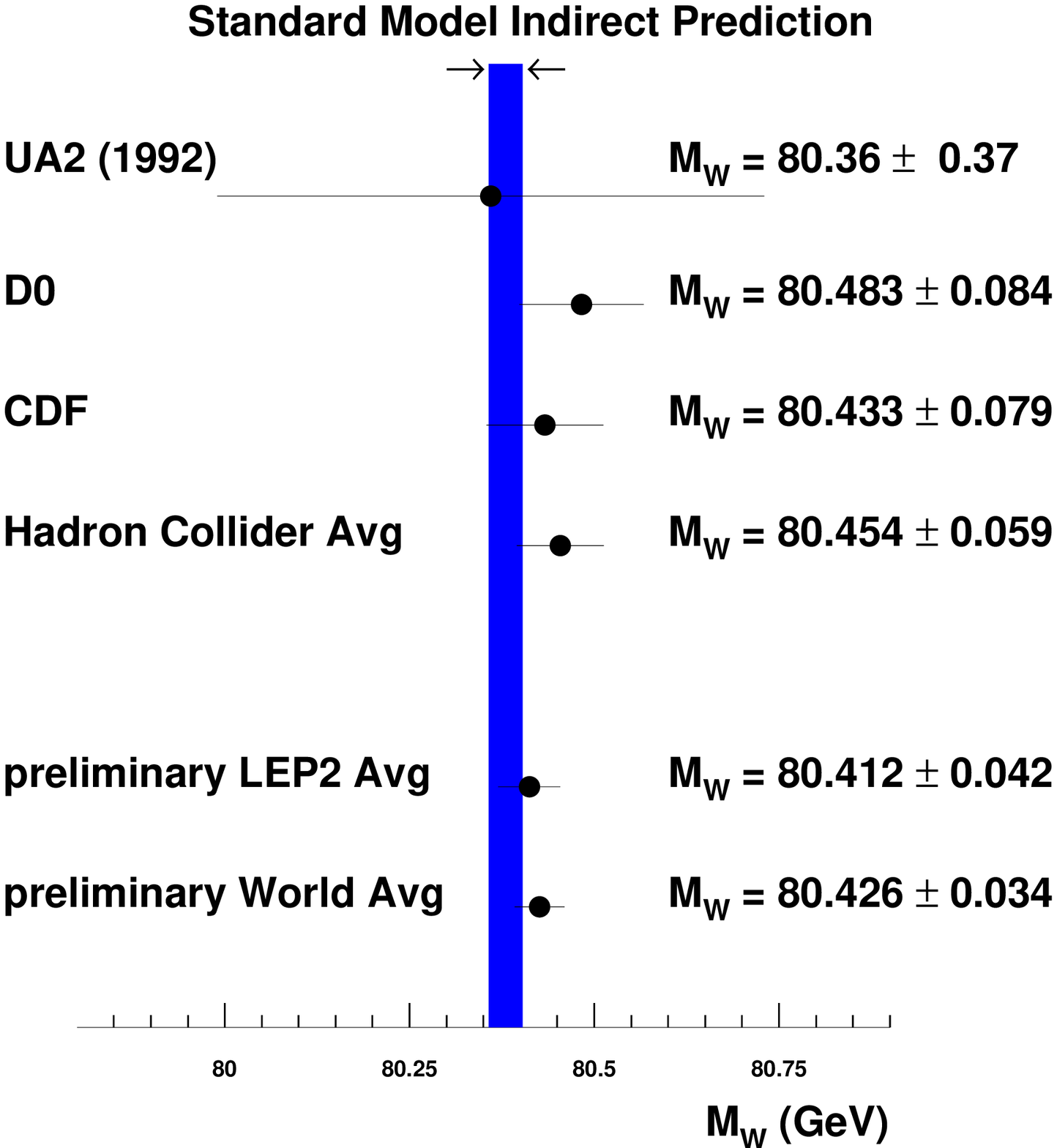,width=7.cm}
\caption{CDF and \dz\ Run 1 combined direct W width (left) and W mass (right) measurements.
\label{fig:run1_combined}}
\end{center}
\end{figure}

The final Tevatron Run~1 CDF and \dz\ combined W~mass and width measurements have been published last year~\cite{run1comb} 
leading to the results shown in Figure~\ref{fig:run1_combined}. 

\subsection{W width measurement}

\begin{figure}[!htb]
\begin{center}
\psfig{figure=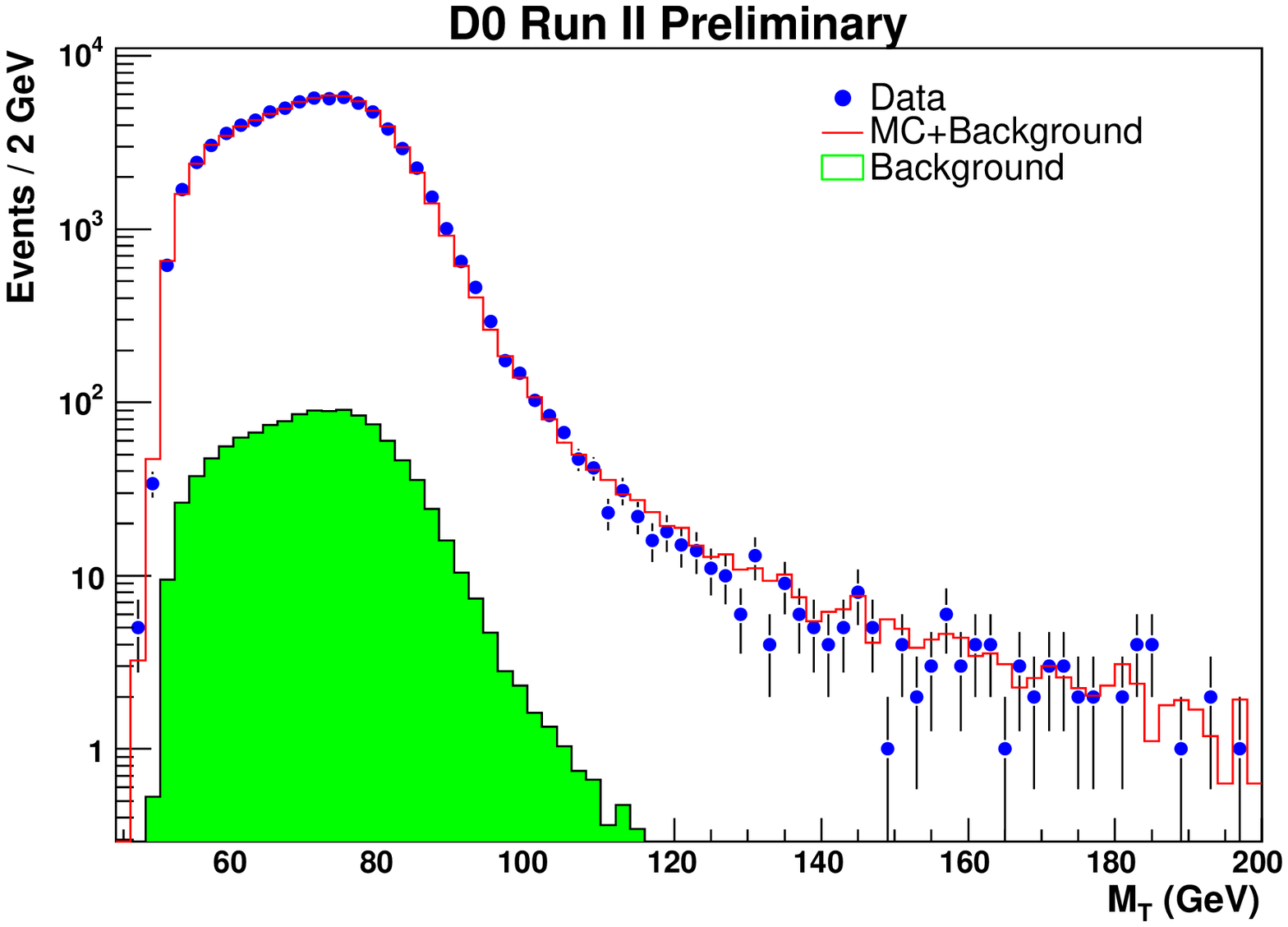,width=7.5cm}
\psfig{figure=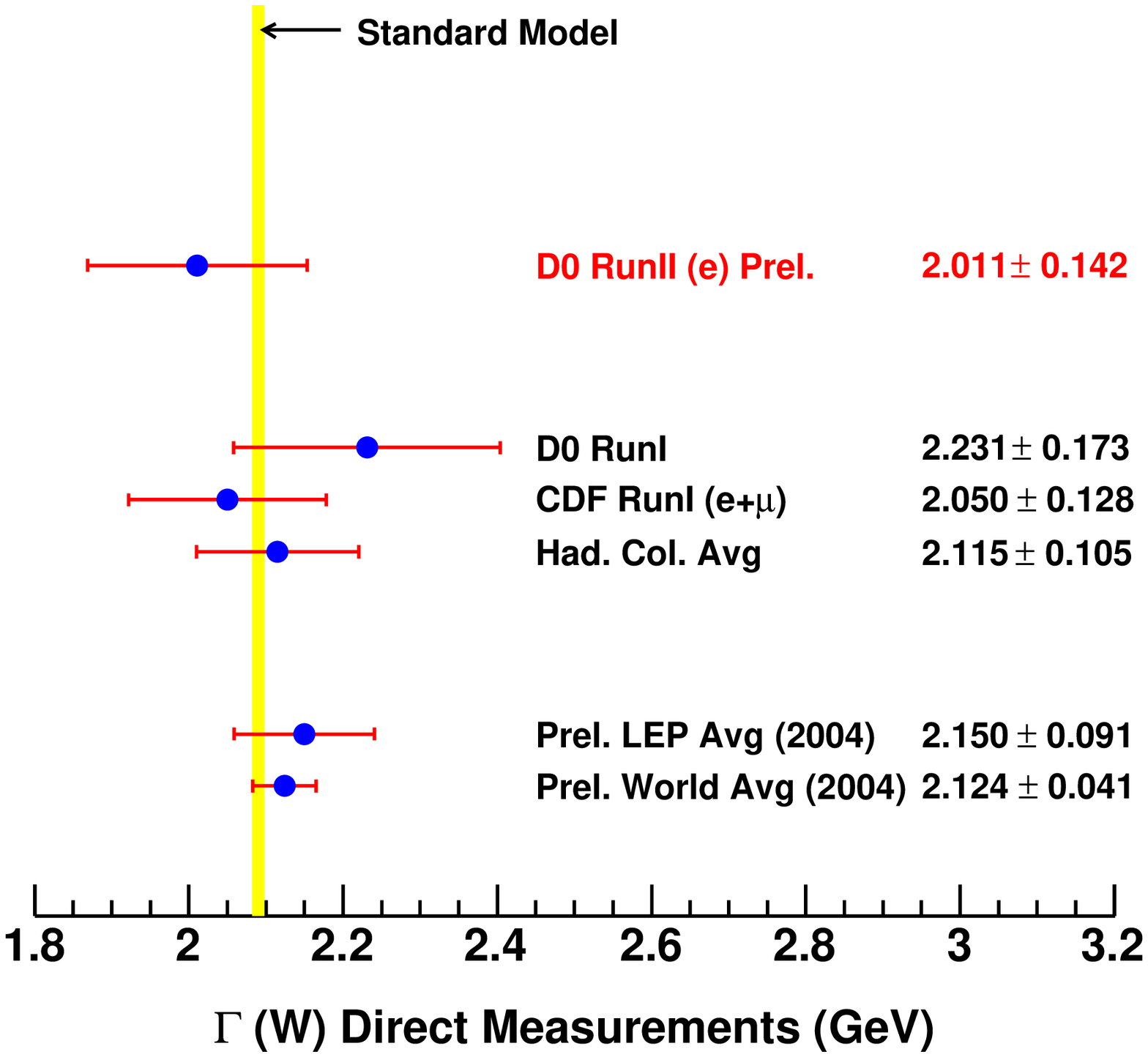,width=7.5cm,height=6cm}
\caption{Left: $W \to e \nu$ transverse mass. The points are the data, the histogram is the Monte Carlo simulation.
 Right: summary of the direct W width measurements.}
\label{fig:direct_wwidth}
\end{center}
\end{figure}

\dz\ has performed a first direct measurement of the W width using $177~\ipb$ of Run~2 data in the electron
channel. The method is to fit the W transverse mass distribution in the region $100 < M_T < 200$~GeV. This 
method requires a simulation program that takes well into account the electromagnetic calorimeter response and 
resolution observed in the data. Figure~\ref{fig:direct_wwidth} (left) shows the data and Monte Carlo agreement.
The main systematic uncertainties in the measurement come from hadronic response and resolution (64~MeV), 
on modelling of the underlying event (47~MeV) and electromagnetic calorimeter response (30~MeV). 
The obtained result is: $$\Gamma_W = 2.011 \pm 0.093 (stat) \pm 0.099 (sys)~\mathrm{GeV.}$$
Figure~\ref{fig:direct_wwidth} shows that the 
achieved uncertainty is comparable to the Run~1 error.

\subsection{W mass analysis}

CDF has performed a first pass of the W~mass analysis using $200~\ipb$ of Run~2 data by determining the uncertainties 
in both the electron and muon channels. For this analysis, the muon momentum scale is set using $J/\psi \to \mu \mu$ and 
$\Upsilon \to \mu \mu$ signals and the electron energy scale is set using the ratio of the energy measured in the calorimeter
with the electron momentum measured in the tracking. This electromagnetic energy scale is corrected for 
non-linearity. The $Z \to ee$ invariant mass is used to tune and cross-check this energy scale. 
The modelling of the hadronic system that recoils against the W boson is tuned using $Z \to \mu \mu$ events.
At large Z transverse momentum $\pt(Z)$, the hadronic resolution is dominated by jet resolution while at low $\pt(Z)$,
it is dominated by the modelling of underlying event. This underlying event is modelled using minimum bias data.

The list of systematic uncertainties in both the electron and muon channels is shown in Table~\ref{tab:wsyst}.
Combining all these uncertainties leads to a global systematic error of 76~MeV which is already lower than
the CDF Run~I systematic error (79~MeV). We expect the first Run~2 W~mass measurement soon.

\begin{table}[!htb]
\caption{Systematic uncertainties on the Run 2 CDF mass determination in MeV.}
\label{tab:wsyst}
\begin{center}
\begin{tabular}{|c|c|c|}
\hline
 & e channel & $\mu$ channel \\
\hline
Lepton energy scale and resolution & 70 & 30 \\
\hline
Recoil scale and resolution & 50 & 50 \\
\hline
Statistics & 45 & 50 \\
\hline
Production and decay model & 30 & 30 \\
\hline
Background & 20 & 20 \\
\hline
\end{tabular}
\end{center}
\end{table}

\section{Conclusion}
The Tevatron has an extensive electroweak program that is well underway. All the inclusive single boson 
cross-sections have been determined including the ones with $\tau$ in the final state.
The \z/Drell-Yan forward-backward asymmetry as well as the \w\ charge asymmetry has been measured 
and the W~mass program is in an advanced stage. All the results found are currently consistent
with the \SM\ expectations.

We expect CDF and \dz\ to provide higher precision measurements in the incoming years as the 
recorded luminosity is growing.

\section*{References}


\begin{thebibliography}{99}
\bibitem{gilles} G. De Lentdecker, {\it Di-boson production at the Tevatron}, These Proceedings.
\bibitem{cdf_xsec} D. Acosta et al. (CDF Collaboration), \PRL~94, 091803 (2005).
\bibitem{ratio_wz_calc} A. D. Martin et al., hep-ph/0308087; \\
P.J. Sutton, A. D, Martin, R.G. Roberts and W. J. Stirling, \PRD 45, 2349 (1992);\\
P. J. Rijken and W. L. van Neerven, \PRD 51, 44 (1995); \\
R. Hamberg, W. L. van Neerven and W. B. Kilgore, \NPB 359, 343 (1991); \\
R. V. Harlander and W. B. Kilgore, \PRL~88, 201801 (2002).
\bibitem{pdg} K. Hagiwara et al., \PRD 66, 010001 (2002).
\bibitem{pdg04} S. Eidelman et al., \PLB 592, 1 (2004).
\bibitem{d0_ztau} V.M. Abazov et al. (\dz\ Collaboration), \PRD 71, 072004 (2005).
\bibitem{SM_xsec} C.R. Hamberg, W. L. van Neerven and T. Matsuura, \NPB 359, 343 (1991).
\bibitem{cdf_afb} D. Acosta et al. (CDF Collaboration), hep-ex/0411059, submitted to \PRD.
\bibitem{cdf_ayw} D. Acosta et al. (CDF Collaboration), \PRD 71, 051104 (2005).
\bibitem{cteq} J. Pumplin, D. R. Stump, J. Huston, H. L. Lai, P. Nadolsky and W. K. Tung, {\em JHEP} 0207:012 (2002).
\bibitem{mrst} A. Martin, R. Roberts, W. Stirling and R. Thorne, {\em Eur. Phys. J.} C 4, 463 (1998).
\bibitem{resbos} F. Landry, R. Brock, P. M. Nadolsky, C. P. Yuan, \PRD 67, 073106 (2003).
\bibitem{lepewwg} LEP Electroweak Working Group, http://lepewwg.web.cern.ch/LEPEWWG/
\bibitem{run1comb} M. Abazov et al. (CDF and \dz\ Collaborations), \PRD 70, 092008 (2004).
\end{thebibliography}
\end{document}